%% file: main.tex
\documentclass{article}
\usepackage[utf8]{inputenc}
\usepackage{graphicx}
\usepackage{geometry}
\usepackage{authblk}
\usepackage{amssymb}
\usepackage{xcolor}
\usepackage{hyperref}
\usepackage{amsmath,amssymb,amsfonts}
\usepackage{soul}
\usepackage{graphicx}
\usepackage{textcomp}
\usepackage{xcolor}
\usepackage[linesnumbered,ruled]{algorithm2e}
\newcommand{\customwidth}{0.49\textwidth}

\usepackage[sorting = none, backend = biber, style=numeric-comp]{biblatex}

\newcommand{\N}{\ensuremath{\mathbb{N}}}
\newcommand{\R}{\ensuremath{\mathbb{R}}}

\title{Advanced unembedding techniques for quantum annealers}

\author[1]{Elijah Pelofske\thanks{Email: epelofske@lanl.gov}}
\author[2]{Georg Hahn\thanks{Email: ghahn@hsph.harvard.edu}}
\author[1]{Hristo Djidjev\thanks{Email: djidjev@lanl.gov}}

\affil[1]{Los Alamos National Laboratory, CCS-3, Los Alamos, NM 87545, USA}
\affil[2]{Harvard University, T.H.\ Chan School of Public Health, Boston, MA 02115, USA}

\date{\vspace{-6ex}}

\begin{document}

\maketitle

\input{abstract}
\input{introduction}
\input{methods}
\input{experiments}

\input{discussion}

\section*{Acknowledgment}
This work has been supported by the US Department of Energy through the Los Alamos National Laboratory. Los Alamos National Laboratory is operated by Triad National Security, LLC, for the National Nuclear Security Administration of U.S.\ Department of Energy (Contract No.~89233218CNA000001) and by the Laboratory Directed Research and Development program of Los Alamos National Laboratory under project numbers 20190065DR and 20180267ER.

\appendix
\input{appendix}

\printbibliography

\end{document}

%% file: abstract.tex
\begin{abstract}
The D-Wave quantum annealers make it possible to obtain high quality solutions of NP-hard problems by mapping a problem in a QUBO (quadratic unconstrained binary optimization) or Ising form to the physical qubit connectivity structure on the D-Wave chip. However, the latter is restricted in that only a fraction of all pairwise couplers between physical qubits exists. Modeling the connectivity structure of a given problem instance thus necessitates the computation of a minor embedding of the variables in the problem specification onto the logical qubits, which consist of several physical qubits "chained" together to act as a logical one. After annealing, it is however not guaranteed that all chained qubits get the same value (-1 or +1 for an Ising model, and 0 or 1 for a QUBO), and several approaches exist to assign a final value to each logical qubit (a process called "unembedding"). In this work, we present tailored unembedding techniques for four important NP-hard problems: the Maximum Clique, Maximum Cut, Minimum Vertex Cover, and Graph Partitioning problems. Our techniques are simple and yet make use of structural properties of the problem being solved. Using Erd\H{o}s-R\'enyi random graphs as inputs, we compare our unembedding techniques to three popular ones (majority vote, random weighting, and minimize energy).We demonstrate that our proposed algorithms outperform the currently available ones in that they yield solutions of better quality, while being computationally equally efficient.
\end{abstract}

\bigskip
\textbf{Keywords:} chained qubits; D-Wave; NP-hard problems; optimization; quantum annealing; unembedding

%% file: introduction.tex
\section{Introduction}
\label{sec:intro}
Quantum annealers of D-Wave Systems, Inc., are designed to find high-quality solutions of problems that can be expressed as the minimization of a function
\begin{align}
    H(x_1,\ldots,x_n) = \sum_{i=1}^n a_i x_i + \sum_{i<j} a_{ij} x_i x_j
    \label{eq:Hamiltonian}
\end{align}
in $n \in \N$ variables, where $a_i \in \R$ and $a_{ij} \in \R$ are parameters specified by the user. The instance in eq.~\eqref{eq:Hamiltonian} is called a \textit{QUBO} (quadratic unconstrained binary optimization) problem, if $x_i \in \{0,1\}$, and an \textit{Ising} problem, if $x_i \in \{-1,+1\}$. Both QUBO and Ising formulations are equivalent as each one can be transformed into the other by a linear transformation of the variables \cite{Chapuis2019}. See \cite{Lucas2014} for a comprehensive overview of many NP-hard problems that can be formulated as QUBO or Ising problems. These include many important problems such as the Maximum Clique \cite{Chapuis2019}, Minimum Vertex Cover \cite{vertex-cover-CF19}, Graph Coloring \cite{TITILOYE2011376}, or Graph Partitioning \cite{ushijima2017graph} problems.

\begin{figure}[h]
    \centering
    \includegraphics[width=0.5\textwidth]{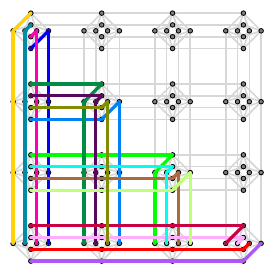}
    \caption{A Chimera graph consisting of a $4\times 4$ array of unit cells (D-Wave 2000Q uses a $16\times 16$ array) and a minor embedding of a clique of size $16$ into it. Variables of the clique are mapped onto chains, which are all here of size 5, and are shown in different colors. Figure taken from \cite{Boothby2020}.\label{fig:chimera}}
\end{figure}

Three steps are required to solve an NP-hard problem on a D-Wave annealer. First, the problem under investigation has to be expressed as minimization of a function of type \eqref{eq:Hamiltonian}. Second, the problem being solved, expressed as minimization of type \eqref{eq:Hamiltonian}, has to be transferred to the D-Wave quantum chip, where coefficients $a_i$ and $a_j$ are encoded in parameters called \textit{biases} of two different qubits, say $q_i$ and $q_j$, and coefficient $a_{ij}$ is encoded as bias of the link connecting $q_i$ and $q_j$. On the D-Wave chip, the physical qubits are arranged in a structure called \textit{Chimera graph}, which, in the case of D-Wave 2000Q, is a lattice of $16 \times 16$ cells each of which is itself a $4 \times 4$ complete bipartite graph, see Figure~\ref{fig:chimera}. Ideally, each unknown $x_i$ in \eqref{eq:Hamiltonian} corresponds to a qubit on the D-Wave architecture. However, the connectivity structure of the physical qubits is rather limited as shown in Figure~\ref{fig:chimera}, whereas in eq.~\eqref{eq:Hamiltonian} arbitrary edges between any pair of qubits are allowed. Therefore, a 1-to-1 mapping preserving all pairwise interactions of the unknowns $x_i$ in eq.~\eqref{eq:Hamiltonian} to the physical qubits on the Chimera graph might not exist. Instead, a \textit{minor embedding} of the theoretical to the physical qubits is usually required, for which several physical qubits are "chained" together to act as one \textit{logical qubit}. This is achieved by assigning negative biases of large magnitude (the magnitude is called the \textit{chain strength}), to the links connecting them. During annealing, the quantum system aims to reach a minimum energy state. Hence, the assigned negative biases on the links encourage the connected qubits to take equal values at the end of the anneal, since this results in a lower energy, assuming the chain strength is large enough. But for reasons we discuss later,  chain strengths are often not sufficiently high, resulting in chains ending up having qubits of different values. Such chains are called \textit{broken chains}. Hence in the third step, the qubits in each broken chain have to all be assigned some final value. This last process is called ``unembedding".

\begin{figure}[h]
    \centering
    \includegraphics[width=0.7\textwidth]{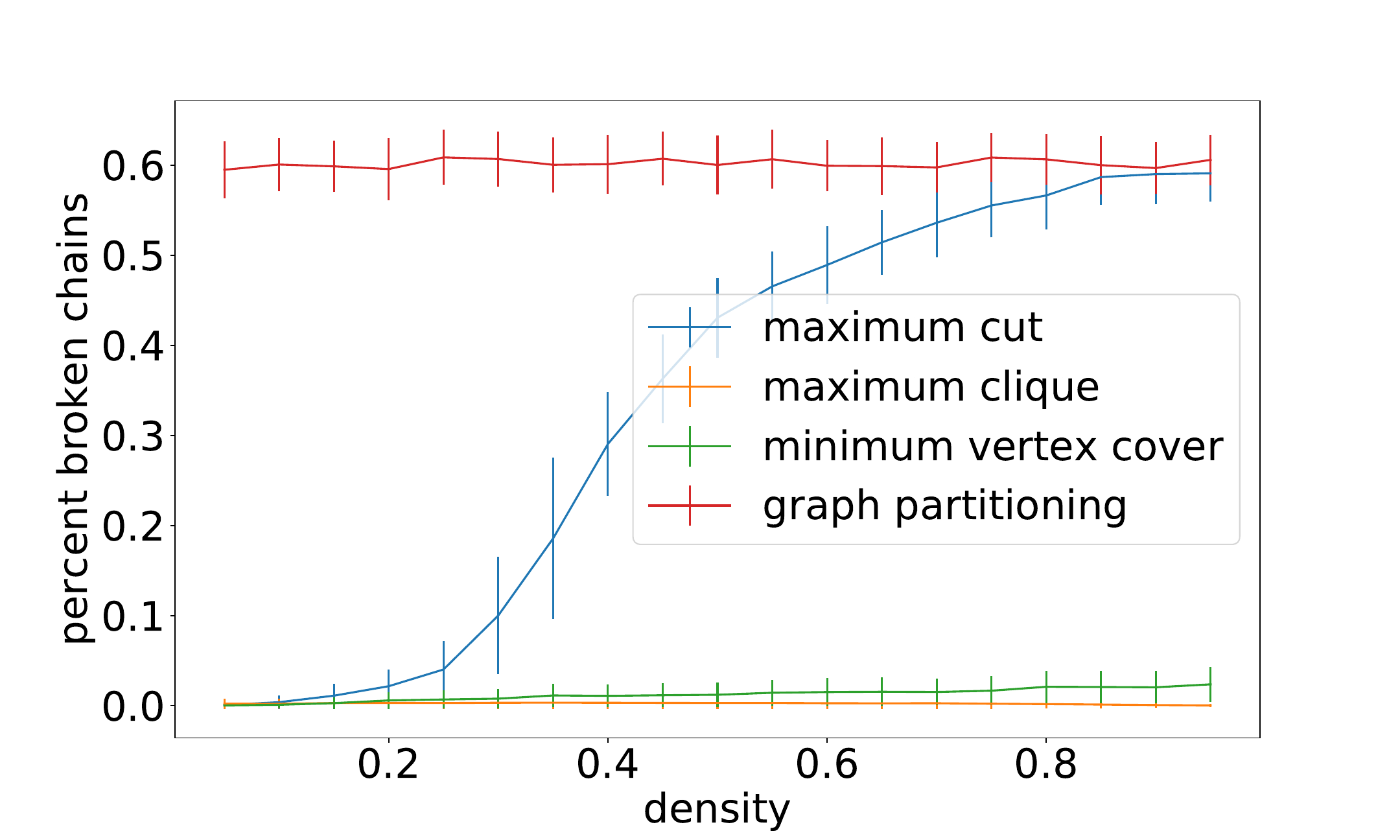}
    \caption{Proportion of broken chains for the Maximum Cut, Maximum Clique, Minimum Vertex Cover, and Graph Partitioning problems for $65$ vertex graphs, as a function of their density. The chain strength is set by using the \texttt{uniform torque compensation} function from the D-Wave SDK, with a default prefactor. Error bars of one standard deviation are given for each data point. Precise parameters of this experiment are given in Section~\ref{sec:experiments}.\label{fig:proportion_brokenchains}}
\end{figure}

Although the computation of a minor embedding allows one to implement arbitrary problems of type \eqref{eq:Hamiltonian} onto the D-Wave architecture, the approach comes with four disadvantages: (a) Computing a minor embedding is itself an NP-hard problem and thus any embedding is computed heuristically. That means that most embeddings are not making optimal use of the Chimera architecture, and some problems of type \eqref{eq:Hamiltonian} might fail to be embedded even if an embedding exists; (b) Chaining qubits together severely reduces the number of available physical qubits, thus restricting the size of the problems being solvable on D-Wave; (c) The chained qubits in an embedding are supposed to represent and act as a single logical qubit. For this a sufficiently large weight on all edges in a chain have to be placed, and it is non trivial to choose an optimal value for the chain weight. This is because, as mentioned above, the choice of the chain weight impacts the problem being solved on D-Wave. If it is too small, chains will not act as a single logical qubit, and if it is too large then the annealer solution might only satisfy the chains but not the actual problem constraints; (d) In order to arrive at a consistent solution for the logical qubits in eq.~\eqref{eq:Hamiltonian} for broken chains, some form of post-processing is required (the unembedding step). The phenomenon of broken chains is very common but depends on the problem being solved: see Figure~\ref{fig:proportion_brokenchains} for the proportion of broken chains occurring for the four graph problems we consider in this article as a function of the graph density. This work focuses on new methods for unembedding chained qubits.

The D-Wave annealer comes with a variety of post-processing techniques for unembedding. The most straightforward one is called ``majority vote". Suppose some logical qubit $x_i$ is mapped onto a chain $x_i^{(1)},\ldots,x_i^{(m)}$ of $m \in \N$ physical qubits. After annealing, the values of $x_i^{(1)},\ldots,x_i^{(m)}$ are read, and $x_i$ is assigned the most common value among the $m$ chained qubits. In case of a tie, $x_i$ is set to $+1$ for both QUBO and Ising models \cite{unembedding}.

Another technique by D-Wave is called "random weighted unembedding". Here, the empirical frequencies of both $0$ and $1$ for a QUBO (and $-1$, $+1$ for an Ising model) in the chain $x_i^{(1)},\ldots,x_i^{(m)}$ are calculated first. The value of $x_i$ is then set to a randomly drawn value $0$ or $1$ (respectively $-1$ or $+1$ for Ising models) with a probability proportional to the empirical frequency.

Last, D-Wave offers to resolve chains with a technique termed "minimize energy", which works as follows. First, the value $v_0$ of the Ising (or QUBO) model restricted to all unbroken chains (defined as those chains whose  qubits all take the same value) is computed. Next, for each chain $i$,  two values $v_i^{(-1)}$ and $v_i^{(+1)}$ of the model are computed, that is, the value of the QUBO/Ising is computed with chain $i$ set to $-1$ or $+1$ added to the set of unbroken chains. Using these three values, a priority score $v_0 - \min \{ v_i^{(-1)}, v_i^{(+1)} \}$ is calculated for each chain $i$. In the main loop of the algorithm, the priorities and their corresponding chain indices are fed into a heap structure and pulled in decreasing order. For each chain under consideration, the chain value is set to $-1$ for Ising (and $0$ for QUBO) models if $v_i^{(-1)}\leq v_i^{(+1)}$. Otherwise, it is set to $+1$ for both Ising and QUBO problems. After having fixed the chain value, all remaining priorities in the heap are updated in the same way as described above. Afterwards, the next chain is pulled from the heap and resolved, and the process continues until no more chains remain.

The aim of this paper is to show that simple linear-time heuristics, tailored to the NP-hard problem being solved on D-Wave, outperform the aforementioned standard techniques for resolving chains in that they yield solutions of better quality, while being computationally equally efficient. Our proposed heuristics conceptionally work in the same way for all problems we investigate, though the precise steps differ depending on the problem: We consider the sets of unbroken chains and broken chains separately. Then, the partial solution spanned by all unbroken chains is taken as a baseline, to which the variables having broken chains are added in such a way that a feasible solution to the original NP-hard problem is guaranteed. This step never requires more time than a linear function of the number of qubits (excluding a possible logarithmic factor), which is a negligible fraction of the total running time and consistent with the times for the three D-Wave algorithms mentioned above.

We evaluate the performance of both the standard D-Wave unembedding algorithms, as well as our techniques, on Erd{\H o}s–R{\'e}nyi random graphs as a function of the graph density. Our experiments demonstrate that the standard D-Wave techniques can be considerably improved upon by using tailored unembedding algorithms such as the ones we present.

The article is structured as follows. After a brief summary of related work in Section~\ref{sec:literature}, Section~\ref{sec:methods} presents the tailored unembedding techniques we propose in this article. These are the unembedding techniques for the Maximum Clique, Minimum Vertex Cover, Graph Partitioning, and Maximum Cut problems. Section~\ref{sec:experiments} contains our experimental analysis. The article concludes with a discussion in Section~\ref{sec:discussion}.

\subsection{Related work}
\label{sec:literature}
While there are only few published results on the unembedding problem, the related embedding problem has been much more researched. The computation of a minor embedding is itself an NP-hard problem \cite{minor-NP}, with the exception of (small) fixed pattern graphs~\cite{Robertson-Seymour}. Therefore, substantial research available in the literature has focused on finding fast and high-quality heuristics. Those heuristics either work by reducing the number of used vertices in a semi-valid existing embedding, by iteratively modifying the placement of a chain to reduce the number of edges, or by using the special structure of the QUBO or Ising model \cite{choi2008minor, date2019efficiently, vyskovcil2019embedding, etezad2016planarity, andreaspaper}.

Another related aspect is how to choose an appropriate chain strength.
In \cite{Venturelli2015}, the authors use a variety of  heuristic formulas to set both the chain strength (see also \cite{Raymond2020}) and the total anneal duration. In \cite{Prudenz2016}, the author considers non-uniform   strategies for assigning biases to chains. In each chain, a bias can be assigned to each physical qubit in the chain that takes into account the number of adjacent active physical qubits and the chain length. Then the method simply sets a logical qubit bias to the value that the single chained qubit with largest weight (as defined above) takes.

On the other side, the unembedding process, that is, the assignment of logical qubits based on the values their chained physical qubits have taken, has largely been neglected in the literature. While we are interested in this paper in simple, single-pass algorithms to resolve chains and arrive at a solution for all logical qubits, other works in the literature have taken more involved routes. For instance, \cite{Ayanzadeh2020} considers reinforcement quantum annealing, in which a classical algorithm evaluates the solution returned by D-Wave and adjusts the penalty of unsatisfied constraints. Other approaches combining quantum annealing with classical simulated annealing have been considered~\cite{Ramezanpour2018}. In \cite{Dorband2018,Borle2019}, an approach is discussed that starts off with a collection of samples from the D-Wave annealer. Using the pool of samples, a new sample is constructed for every pair of samples, with the property that its energy is never higher than the one of the two samples it was derived from. This process is repeated until only one sample remains.

All these approaches, in essence, employ an entire hybrid quantum-classical algorithm, where the classical part is used to either iteratively call the quantum annealer to improve the solution from the previous iteration, or to run a fully-fledged classical optimization algorithm after getting the results from the quantum annealer, often not only computing values for the broken chains, but also modifying the already computed values of the unbroken ones. In contrast, we only unembed broken chains, and do not allow changing the value of a logical variable computed by the annealer, consistent with the unembedding algorithms implemented by D-Wave.

%% file: methods.tex
\section{Tailored unembedding techniques}
\label{sec:methods}
This section presents the tailored unembedding algorithms that we propose for the Maximum Clique, Maximum Cut, Minimum Vertex Cover, and Graph Partitioning problems. Those problems are defined as follows for an arbitrary graph $G=(V,E)$: (1) A cut of $G$ is a partition of $V$ into two disjoint subsets $S$ and $T$ satisfying $V = S \cup T$. The size of the cut given by $S$ and $T$ is the number of \textit{cut edges} between vertices of the two sets. The Maximum Cut problem asks for the cut having the largest cut size. (2) A \textit{clique} $C \subseteq V$ is a complete subgraph. The Maximum Clique problem asks for clique of maximum size. (3) A \textit{vertex cover} $C \subseteq V$ is subset such that all edges in $E$ have at least one endpoint at a vertex contained in $C$. The Minimum Vertex Cover problem asks for the vertex cover of minimum size. (4) The Graph Partitioning problem asks for a partition of $V$ into two subsets $P_1$ and $P_2$ (satisfying $V = P_1 \cup P_2$ and $|P_1\cap P_2=\emptyset$), which is balanced, i.e. $|\,|P_1|-|P_2|\,|\leq 1$, and such that the number of edges between $P_1$ and $P_2$ is minimized.

The algorithms themselves differ for the four NP-hard problems we consider, however their conceptional idea is consistent. Briefly, all methods begin by determining the set $U$ of all unbroken chains and the set $B$ of all broken chains. The solution spanned by the unbroken chains having value $1$ is then used as a baseline solution (e.g., an initial clique or a partial partitioning). The main part of all algorithms consists in iterating over the broken chains in set $B$. During this iteration, context specific knowledge of the problem being solved is used to determine what value the variable corresponding to the broken chain should be assigned.

We may be able to increase the efficacy of any algorithm if we use additional information provided by the annealer. Instead of treating all broken chains equally, we also take into account the percentage of $1$, as well as the percentage of $0$ (or $-1$) assigned on the qubits of each chain. We use this in a case of a tie, when, e.g., we have to assign to one of a pair of vertices a value of 1, we will prefer the vertex that has a higher proportions of 1s in its corresponding broken chain.

\subsection{Maximum Clique unembedding}
\begin{algorithm}[t]
    \caption{Maximum Clique unembedding\label{algo:maxclique}}
    \SetKwInOut{Input}{input}
    \SetKwInOut{Output}{output}
    \Input{set of chains $S$\;}
    \Output{clique $C$\;}
    Determine the set $U$ of unbroken chains and the set $B$ of broken chains, i.e.,\ $U \cup B = S$\;
    Let $U'$ be all variables corresponding to chains in $U$ that are assigned $1$\;
    Check if $U'$ forms a clique and, if not, \textbf{return} an empty clique\;
    \While{$B\neq \emptyset$}{
        Define $L=\{x\in B \;|\; x$ is adjacent to all vertices of $U$\}\; 
        If $L=\emptyset$ then break the loop\;
        Let $L_{max}\subseteq L$ be the set of all nodes whose degree in the subgraph spanned by $L$ is maximum\; 
        Choose $x$ as a  vertex in $L_{max}$ whose chain maximizes the proportion of 1 values relative to its length\;
        Remove $x$ from $B$ and add it to $U$
    }
    Let $C$ be the subgraph of $G$ induced by $U$\;
    \Return{$C$}\;
\end{algorithm}
The pseudo-code of our proposed algorithm for unembed chains for the Maximum Clique problem is given in Algorithm~\ref{algo:maxclique}. Briefly, the algorithm works in two steps: First, all unbroken chains are collected in $U$ and it is checked if those with value $1$ (collected in a set $C$) form a clique. If they do not, a clique of size zero is returned and the algorithms stops (this also applies to the case when the vertices of value $1$ in $U$ do not form a clique but contain a clique).

Otherwise, we try to augment $U$ with vertices and edges  from the subgraph induced by $B$ to get a larger clique. For this, we consider the set $L$ of all vertices in $B$ that are adjacent to all the vertices from $U$. It is easy to see that if any vertex from $L$ is added to the vertices of $U$, the resulting set of vertices defines a (larger) clique in $G$. Hence, we pick such a vertex in $L$, but if there is more than one choice, we pick one with a maximal degree, since such a heuristic is known to increase the chance of ending up with a clique of larger size. If this choice is unique, we add the vertex to the clique. If more than one such vertices of maximum degree exist, we break ther ties by adding  a vertex that has a highest proportion of values $1$ in its chain. We repeat until the current clique can no longer be extended.

\subsection{Maximum Cut unembedding}
\begin{algorithm}[t]
    \caption{Maximum Cut unembedding\label{algo:maxcut}}
    \SetKwInOut{Input}{input}
    \SetKwInOut{Output}{output}
    \Input{set of chains $S$\;}
    \Output{cut $C$\;}
    Determine set $U$ of unbroken chains and set $B$ of broken chains, i.e.\ $U \cup B = S$\;
    Consider cut $C$ (the $-1$ partition and the $+1$ partition) spanned by $U$\;
    \For{$x \in B$}{
        Find the degree of the vertex corresponding to chain $x$ in both partitions of $C$\;
        Allocate $x$ to the partition in $C$ in which it has the lower degree\;
        In case of a tie, choose the partition depending on whether the chain $x$ has more $-1$ or $+1$ values. If no unique choice exists, allocate at random\;
    }
    \Return{$C$}\;
\end{algorithm}
Conceptually, the unembedding algorithm for the Maximum Cut problem (Algorithm~\ref{algo:maxcut}) is the most simple one. We again use the information provided by unbroken chains to inform an initial partial cut. To be precise, the two partitions of the cut are given by the variables having unbroken chains of value $-1$, and of value $+1$ (for an Ising model). We then iterate through all broken chains in arbitrary order, find the degree of the vertex corresponding to a broken chain in both sides of the cut, and allocate it to the side in which it has the lower degree. As before, in case of a tie, we consider the proportion of zeros and ones in the chain and allocate the vertex to the side of higher proportion. If this criterion also fails, we allocate at random.

Algorithm~\ref{algo:maxcut} appears similar to the minimize energy technique of D-Wave, however there are two crucial differences. Minimize energy looks at all vertices $x \in B$, computes how their unembedding will affect the energy, and then chooses the $x \in B$ and the partition to allocate it to such that the energy reduction as greatest. In our implementation of Algorithm~\ref{algo:maxcut}, we look at the vertices $x \in B$ in random order, and only then decide which partition $x$ is added to. The other difference is that we use annealing results to break ties in step 6 of Algorithm~\ref{algo:maxcut}.

\subsection{Graph Partitioning unembedding}
\begin{algorithm}[t]
    \caption{Graph Partitioning unembedding\label{algo:gp}}
    \SetKwInOut{Input}{input}
    \SetKwInOut{Output}{output}
    \Input{set of chains $S$\;}
    \Output{partitioning $P$\;}
    Determine set $U$ of unbroken chains and set $B$ of broken chains, i.e.\ $U \cup B = S$\;
    Consider partitioning $P$ (the $-1$ partition and $+1$ partition) spanned by $U$\;
    \While{both partitions in $P$ contain $\leq \lfloor \frac{|V|}{2} \rfloor$ nodes}{
        Pick random vertex $v$ belonging to a broken chain in $U$\;
        Find degree of $v$ in both partitions of $P$ and allocate $v$ to the partition in which it has the smaller degree\;
        In case of a tie, choose the partition depending on whether the chain of $v$ has more $-1$ or $+1$ values. If no unique choice exists, allocate to the smaller partition\;
    }
    Assign all remaining vertices belonging to chains in $B$ to the partition of size less than $\lfloor \frac{|V|}{2} \rfloor$\;
    \Return{$P$}\;
\end{algorithm}
Our unembedding technique for the Graph Partitioning problem, given in Algorithm~\ref{algo:gp}, is similar to the one for the Maximum Cut problem. However, we have to take into account that the allocation of vertices to the two partitions of the graph stays balanced.

To achieve this, we first proceed as in the case of the Maximum Cut problem. We use the unbroken chains of value $-1$ and $+1$ to inform an initial partial partitioning. We then loop over all broken chains with the aim to allocate them to any of the two partitions. As in this work we consider unembedding from the complete graph of $65$ vertices from the D-Wave 2000Q Chimera graph, the loop in Algorithm~\ref{algo:gp} stops as soon as one partition has reached $32$ nodes.

Inside the loop, we again pick an arbitrary vertex $v$ belonging to a broken chain, find its degree in both existing partitions, and allocate it to the partition in which it has the smaller degree, since in this way we keep the number of edges between partitions low. In case of a tie, we resort to the chain information and choose the $-1$ partition or the $+1$ partition depending on whether the chain of $v$ has more $-1$ or $+1$ values. If this criterion fails, we allocate $v$ to the partition that is smaller since this will more likely result in a balanced partitioning after termination of the algorithm.

\subsection{Vertex Cover unembedding}
\begin{algorithm}[t]
    \caption{Vertex Cover unembedding\label{algo:vertexcover}}
    \SetKwInOut{Input}{input}
    \SetKwInOut{Output}{output}
    \Input{set of chains $S$\;}
    \Output{vertex cover $C$\;}
    Determine set $U$ of unbroken chains and set $B$ of broken chains, i.e.\ $U \cup B = S$\;
    Let $C$ be the set of variables having unbroken chains in $U$ of value $1$, and $Z$ be the ones in $U$ having value $0$\;
    Let $V$ be the set of nodes corresponding to broken chains in $B$\;
    If for any $v \in Z$ we have $N_v \cap Z \neq \emptyset$,  return a trivial cover of all vertices, where $N_v$ is the set of neighbors of $v$ in $G$\;
    Let $X$ be the set of broken chain neighbors of any node in $Z$\;
    Add all vertices in $X$ to $C$\;
    Remove $X$ from $V$\;
    \While{$V \neq \emptyset$}{
        Determine node $v \in V$ having the highest value of deg$_V(v)+f_v$, where deg$_V(v)$ is the degree of $v$ in the subgraph induced by $V$, and $f_v$ is the proportion of ones in the chain\;
        If $Z \cap N_v \neq \emptyset$ then add $v$ to $C$, otherwise add $v$ to $Z$\;
        Remove $v$ from $V$\;
    }
    \Return{$C$}\;
\end{algorithm}
Algorithm~\ref{algo:vertexcover} details the pseudo-code of the unembedding technique we propose for the Vertex Cover problem. Here, we again first look at all unbroken chains in the set $U$, particularly those of value zero (called set $Z$) and those of value one (called the core $C$). The core $C$ is the basis of the cover. We also let $V$ be the set of all vertices corresponding to broken chains in $B$.

Next, we check in line~4 if there is an edge $e$ whose both endpoints have already been assigned to zero (set $Z$). If this is the case, then there is no extension of $C$ (unembedding) that is a vertex cover (we would not be able to cover $e$), and we thus return the trivial vertex cover consisting of all vertices.

Otherwise, we proceed by looking at the neighbors of all nodes in $Z$, collected in a set $X$. Since, by definition, all vertices in $X$ connect to a node in $Z$ assigned zero (i.e., not belonging to the vertex cover), they must themselves be in the vertex cover. Hence, we add all vertices of $X$ to the core $C$ and remove them from $V$.

If $V$ is not empty, we iteratively remove vertices from $V$ according to the priorities  assigned to them and put them either to the set $C$ or to the set $Z$. The priority of a vertex $v$ is defined as the sum of the degree of $v$ in the subgraph of $G$ spanned by $V$ and the proportion of values 1 in the chain corresponding to $v$. The rationale behind this is that such a vertex is likely to cover many others unconsidered vertices in $V$. We again check if $v$ has neighbors that are assigned zero (i.e., which are in set $Z$). If so, that means that $v$ has to be assigned 1  and we add $v$ to the cover $C$. Otherwise, we add $v$ to $Z$. In both cases, we remove $v$ from $V$. The last state of the core $C$ is returned as output of the algorithm.

%% file: experiments.tex
\section{Experiments}
\label{sec:experiments}
This section presents our experiments on the unembedding methods we propose. The quality measure is the difference in the objective value, e.g., the difference in the clique size, cut size, vertex cover size, or partitioning cut, depending on the problem, after unembedding, found by our algorithm and the three default unembedding algorithms of D-Wave Systems outlined in Section~\ref{sec:intro} (majority vote, random weighted, and minimize energy).

All measurements we report are averages over $20$ Erd\H{o}s-R\'enyi random graphs of size $65$ vertices, the maximal size embedabble on the D-Wave Chimera architecture. The annealing time is always $1000$ microseconds, and the number of anneals is $1000$. The chain strength was always determined with the \textit{uniform torque compensation} functionality provided by D-Wave. In Appendix~\ref{sec:experiments_chainstrength} we show that the performance of our algorithms can be improved further by choosing smaller chain strengths.

The Maximum Clique and the Minimum Vertex Cover problem have QUBO formulations (that is, all variables take values in $\{0,1\}$). For a graph $G=(V,E)$, see Section~\ref{sec:methods}, the QUBO of the Maximum Clique problem is given by
$$H=-\sum_{v \in V} x_v + 2 \sum_{(u,v) \in E} x_u x_v,$$
see \cite{Pelofske2019}. For the Minimum Vertex Cover problem, the QUBO is given by
$$H=\sum_{v \in V} x_v + 2 \sum_{(u,v) \in E} (1-x_u)(1-x_v),$$
see \cite{Lucas2014}. The other two problems we look at, the Maximum Cut and Graph Partitioning problems, have Ising formulations (that is, all variables take values in $\{-1,+1\}$). For Maximum Cut, the Ising model is given by
$$H=\sum_{(u,v) \in E} x_u x_v,$$
see \cite{Hahn2017ReducingBQ}. For Graph Partitioning, the Ising model is given by
$$H=A \left( \sum_{v \in V} x_v \right)^2 + \sum_{(u,v) \in E} \frac{1-x_u x_v}{2},$$
where $A$ can be chosen as $\min(|V|,\Delta)/8$, with $\Delta$ being the maximum degree of $G$, see \cite{Lucas2014}. Note that in their original formulations, the Maximum Clique and Maximum Cut problems are maximization problems, whereas for Minimum Vertex Cover and Graph Partitioning the objective is minimization.

In the plots we report the improvement of our method over each of D-Wave's default unembedding methods. For minimization problems, we define the improvement as the quotient of the (vertex cover size or cut size) returned by D-Wave's unembedding divided by the one returned by our unembedding methods. For maximization problems, we define the improvement as the quotient of the (clique size or cut size) returned by our unembedding methods divided by the one returned by DWave's unembedding. That way, for each problem, the more positive a measurement is on the y-axis, the more our custom method outperformed the particular D-Wave unembedding method.

\begin{figure*}[h]
    \centering
    \includegraphics[width=\customwidth]{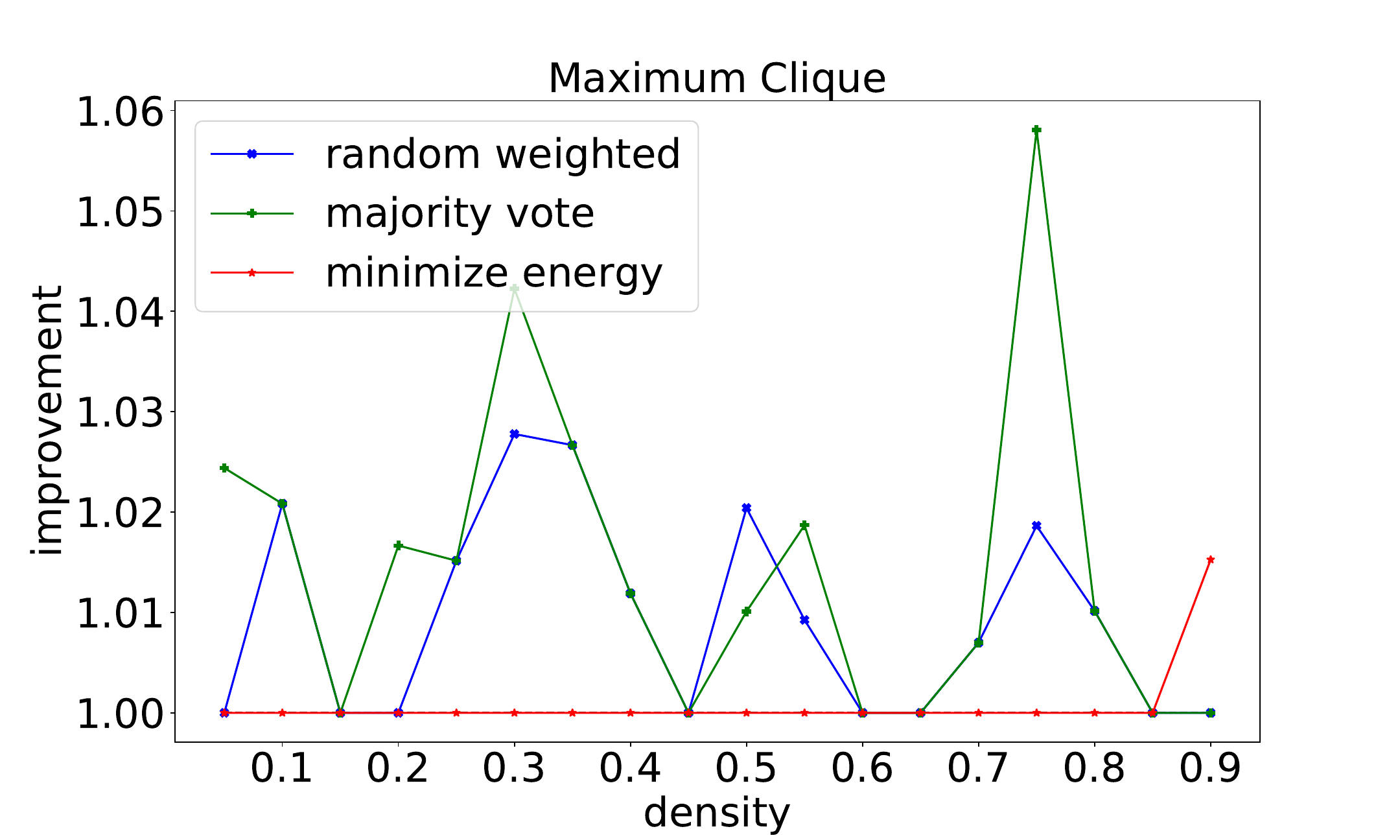}\hfill
    \includegraphics[width=\customwidth]{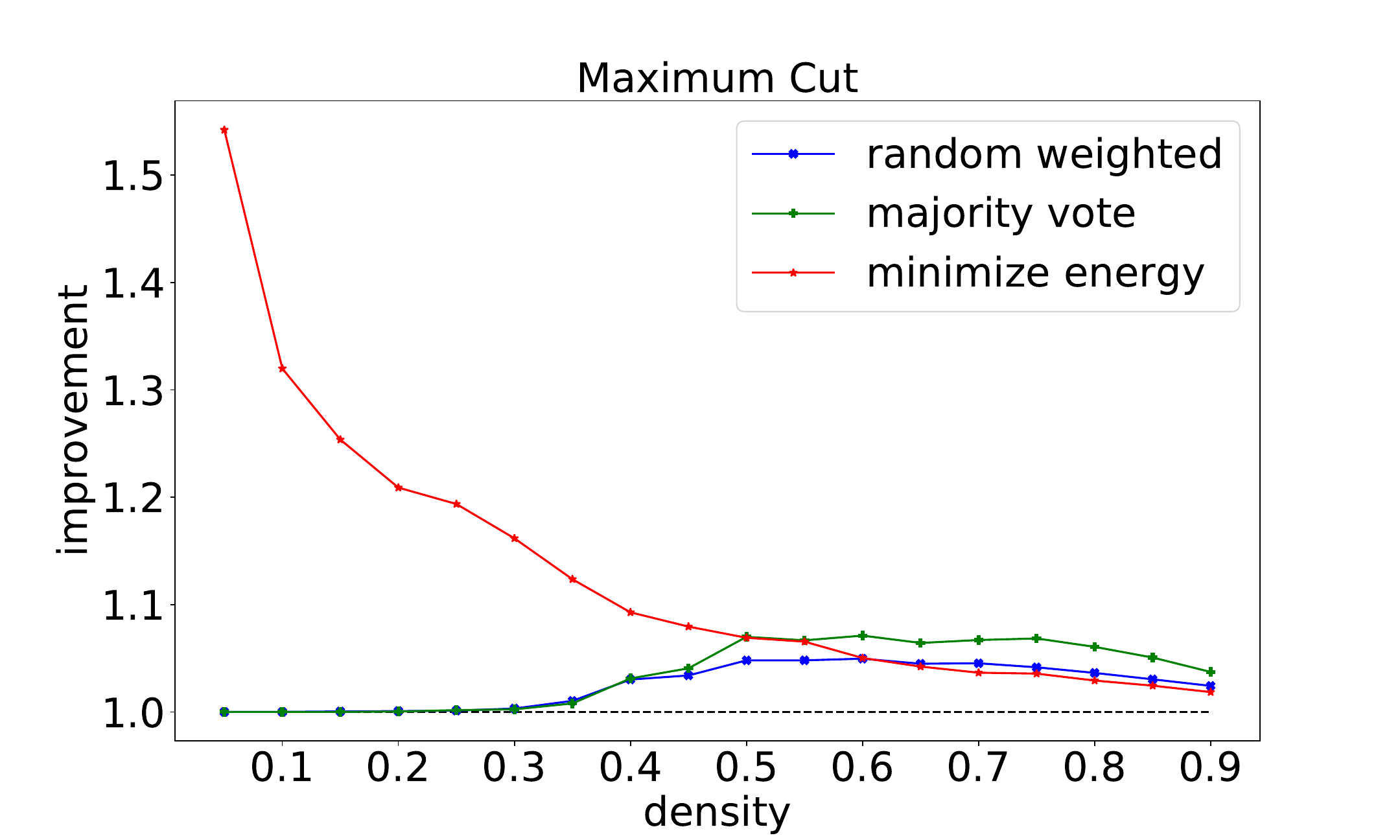}\\
    \includegraphics[width=\customwidth]{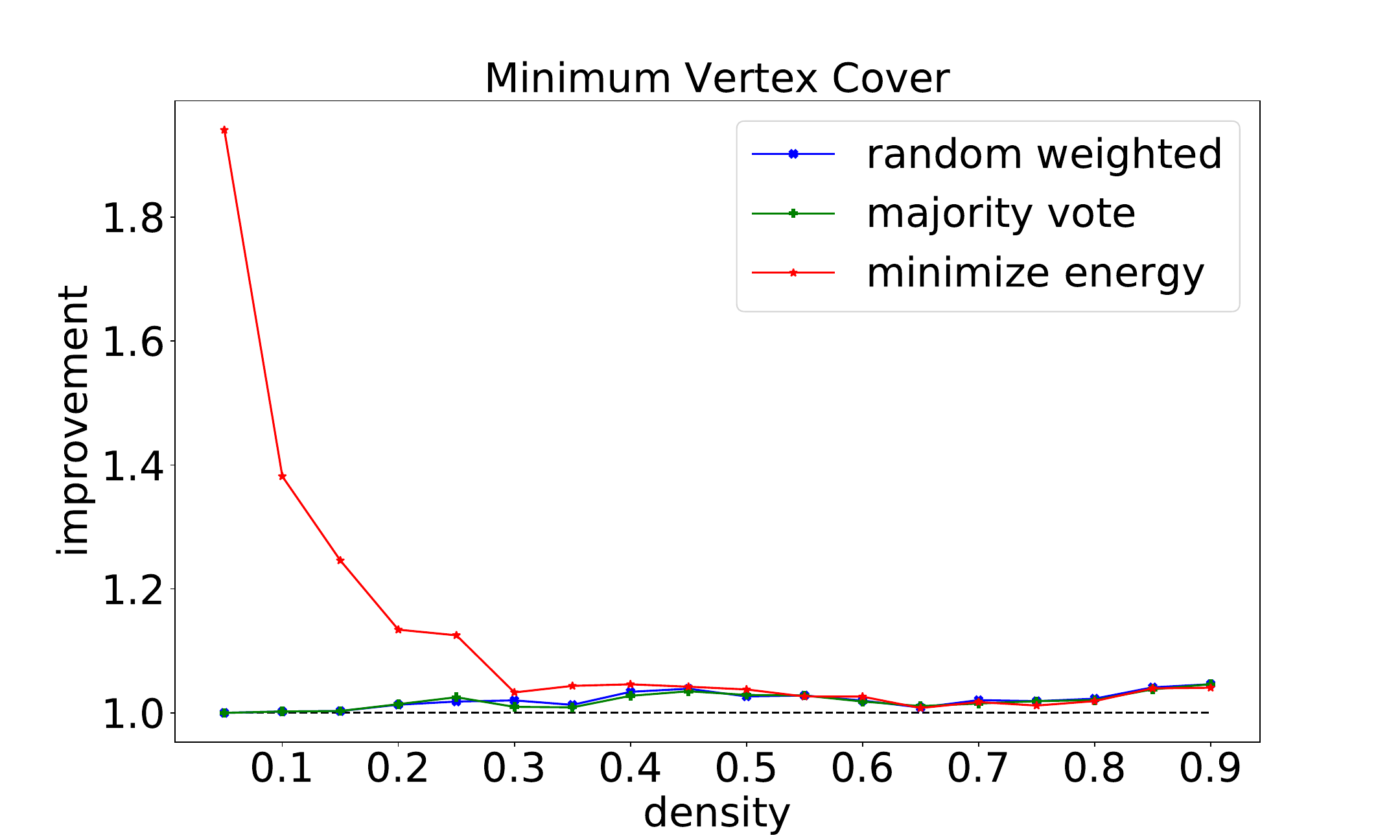}\hfill
    \includegraphics[width=\customwidth]{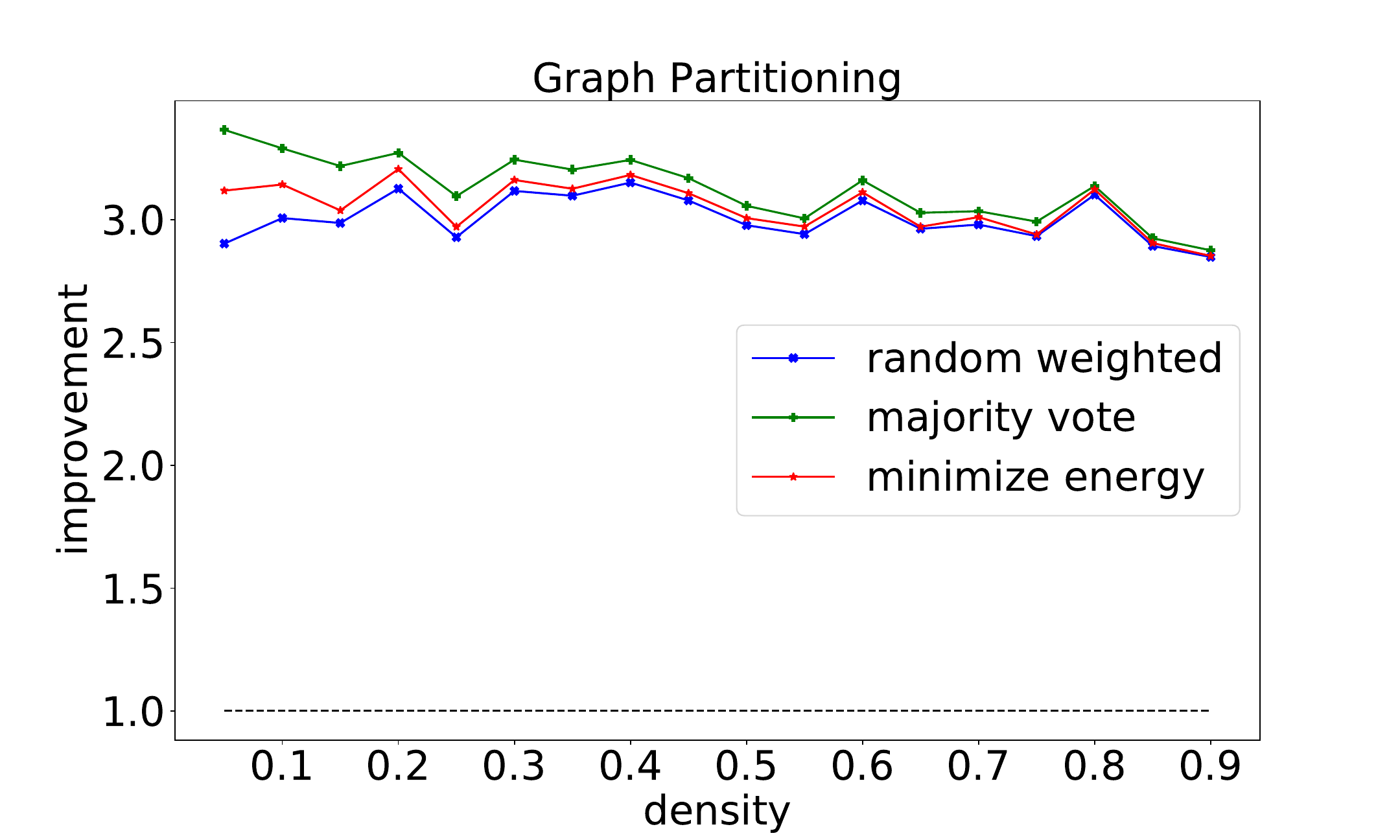}
    \caption{Evaluation of our unembedding techniques as a function of the graph density. Benchmark are the three default unembedding options (majority vote, random weighted, minimize energy) provided by D-Wave Systems, Inc. The improvement measure is defined as the quotient of the (clique size or cut size) returned by our unembedding methods divided by the one returned by DWave's unembedding for the Maximum Clique (top left) and Maximum Cut (top right) problems, respectively. For Minimum Vertex Cover (bottom left) and Graph Partitioning (bottom right), the improvement is defined as the quotient of the (vertex cover size or cut size) returned by D-Wave's unembedding divided by the one returned by our unembedding methods.\label{fig:unembedding}}
\end{figure*}
We proceed to evaluate our proposed algorithms with respect to the three default techniques provided by D-Wave Systems, Inc. Those are majority vote, random weighted unembedding, and minimize energy (see Section~\ref{sec:intro}).

Figure~\ref{fig:unembedding} presents results for all four problems under investigation, where the y-axis displays the improvement factor (in clique size, cut size, vertex cover size) of our methods compared to the three D-Wave default methods. We observe that for the Maximum Clique problem (top left), we draw equal with the "minimize energy" option of D-Wave, while all other default methods of D-Wave are slightly worse across all density scenarios.

For the Maximum Cut problem (top right), the behavior is different. Here, the minimize energy approach of D-Wave performs substantially worse than all other methods (including ours) for low densities, while we draw equal with the "random weighted" and "majority vote" techniques. For higher densities, minimize energy draws equal with the other two default D-Wave techniques, though all of them perform slightly worse than our approach of Algorithm~\ref{algo:maxcut}.

The results for the Minimum Vertex Cover problem (bottom left) are qualitatively similar to the ones for the Maximum Cut problem. Again the minimize energy option of D-Wave performs worse than the other methods for low densities, although it improves for higher densities. The other two default methods draw roughly equal with our approach in Algorithm~\ref{algo:vertexcover}.

An interesting picture is seen for the Graph Partitioning problem (bottom right). Here, all three default methods provided by D-Wave seem to be substantially worse than our unembedding approach of Algorithm~\ref{algo:gp} (by a factor of roughly $3$) for all densities considered.

Though we observe that some methods draw equal for certain ranges of the graph density, in none of the experiments did any of the D-Wave algorithms outperform ours.

%% file: discussion.tex
\section{Discussion}
\label{sec:discussion}
Solving an NP-hard problem on a D-Wave quantum annealer necessitates the computation of an embedding of the logical structure of its Ising or QUBO representation onto the structure (architecture) of the physical qubits on the D-Wave chip. In this process, several qubits are "chained" together to act as one theoretical qubit. The problem with this approach consists in the fact that chained qubits are supposed to all take the same (consistent) value ($0$ or $1$ for QUBO, and $-1$ and $+1$ for Ising models) after reading out the annealing result, though this is not guaranteed in practice. For this reason, D-Wave provides several "unembedding" techniques to assign a final value to each logical qubit.

In this work we show that the default approaches provided by D-Wave can be considerably improved upon by using tailored unembedding techniques for the problem being solved. Our algorithms are easy to implement and computationally equally efficient as the ones of D-Wave. Nevertheless, we observe that our algorithms (almost) uniformly improve upon the ones of D-Wave for Erd\H{o}s-R\'enyi random graphs across all graph densities considered.

This work leaves considerable scope for future work. First and foremost, the presented unembedding techniques can be refined with the aim to improve the quality of the unembedding. Also, since all presented algorithms are problem specific, new algorithms for additional important NP-hard problems such as Graph Coloring, SAT (Satisfiability), etc.\ need to be developed. Most importantly, it would be of interest to frame all developed unembedding techniques in a unified framework, from which unembedding algorithms for a variety of problem classes would follow instantly.

%% file: appendix.tex
\section{Dependence on the choice of the chain strength}
\label{sec:experiments_chainstrength}
In the experiments of Section~\ref{sec:experiments} we used an individual chain strength determined with the \textit{uniform torque compensation} functionality provided by D-Wave. However, lower (higher) chain strengths typically cause more (less) broken chains, thus impacting the performance of our unembedding algorithms (as well as the ones of D-Wave). We are thus interested in evaluating the performance of our techniques as a function of the chain strength.

As in Section~\ref{sec:experiments}, we fix the number of vertices of the random graphs at $65$, and regard three graph densities $p \in \{0.1,0.5,0.9\}$. We embed random instances of the Maximum Clique, Maximum Cut, Minimum Vertex Cover, and Graph Partitioning problem, for each density $p$, onto the D-Wave architecture and unembed the chains using our algorithms of Section~\ref{sec:methods}. We always depict on the y-axis the clique, cut, or vertex cover size after unembedding.

\begin{figure*}[h]
    \centering
    \includegraphics[width=0.49\textwidth]{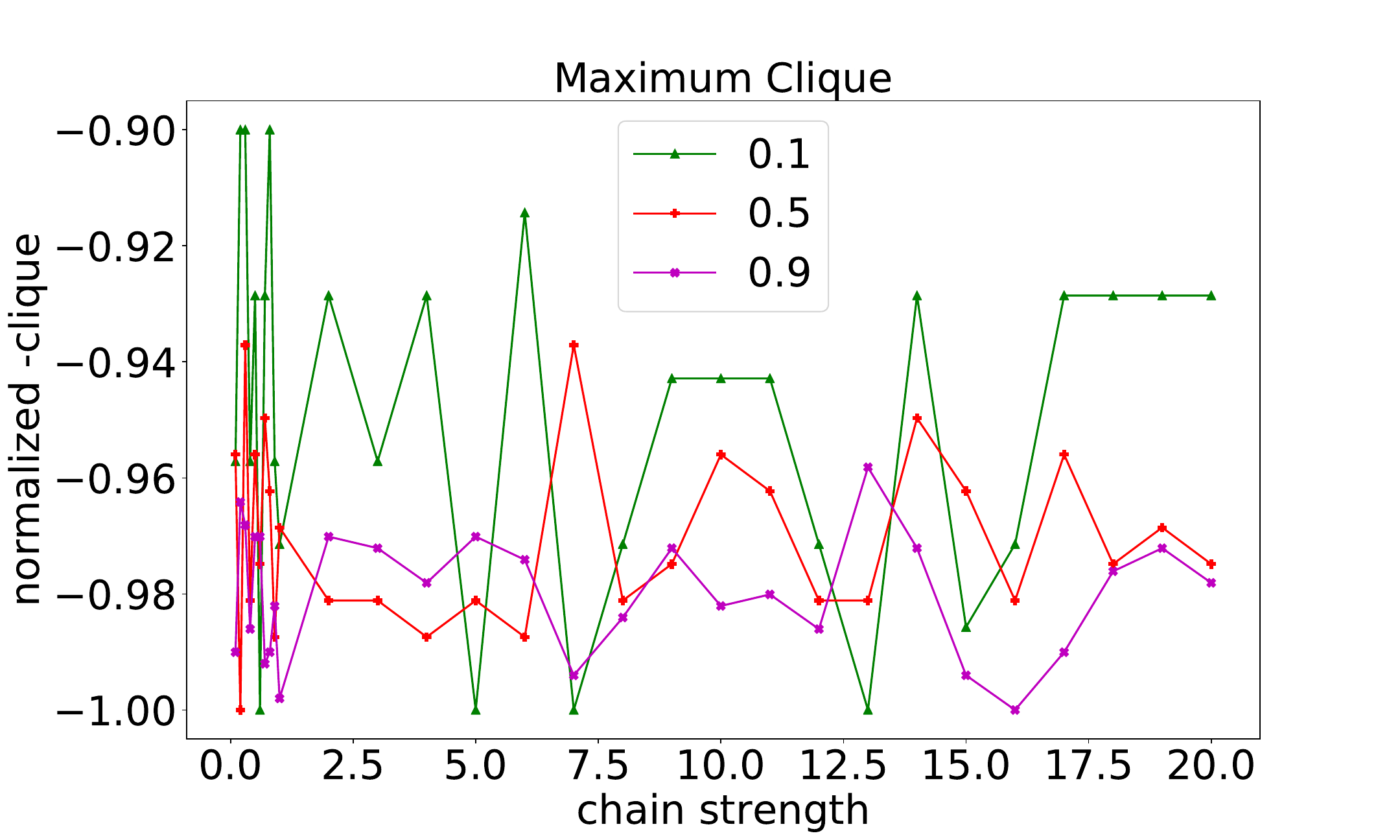}\hfill
    \includegraphics[width=0.49\textwidth]{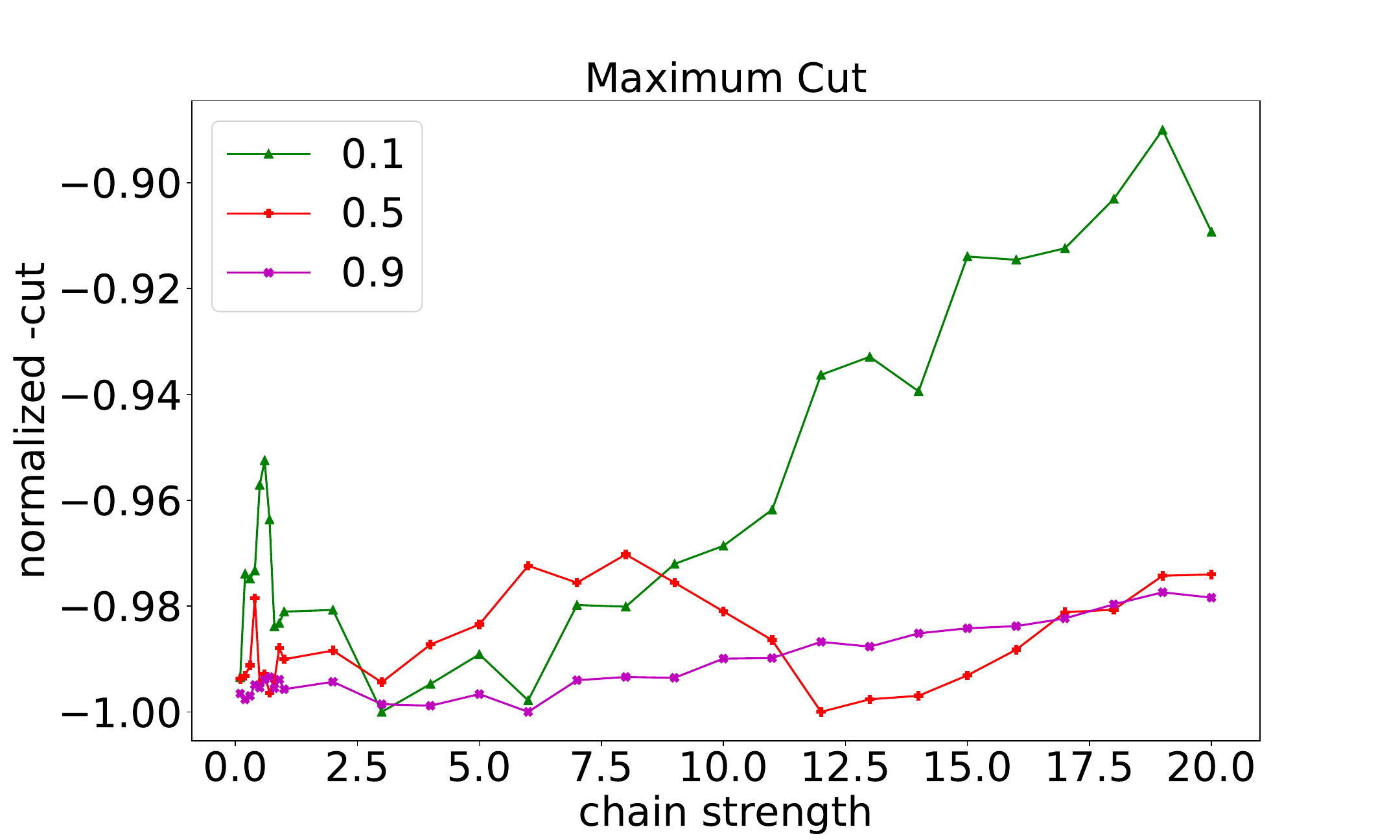}\\
    \includegraphics[width=0.49\textwidth]{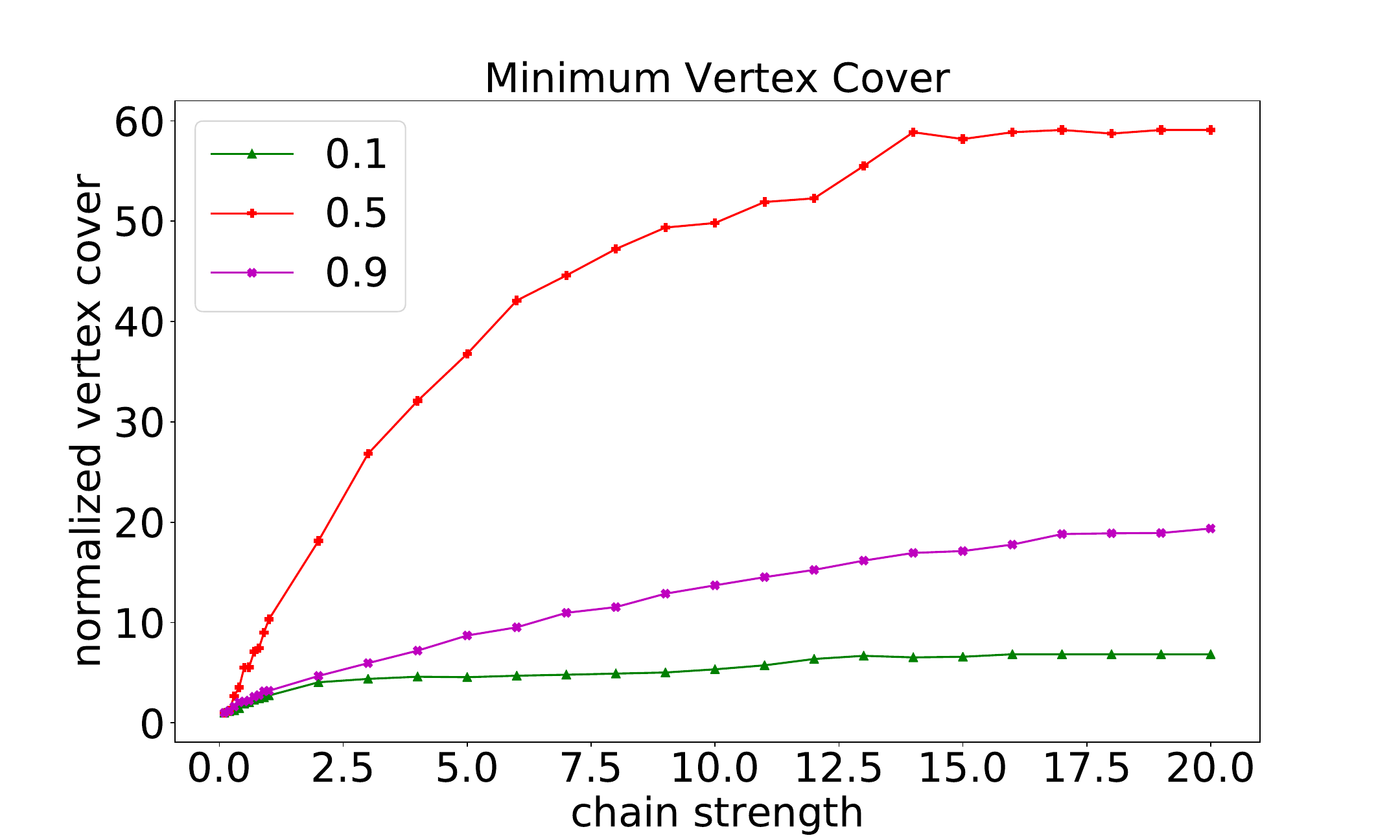}\hfill
    \includegraphics[width=0.49\textwidth]{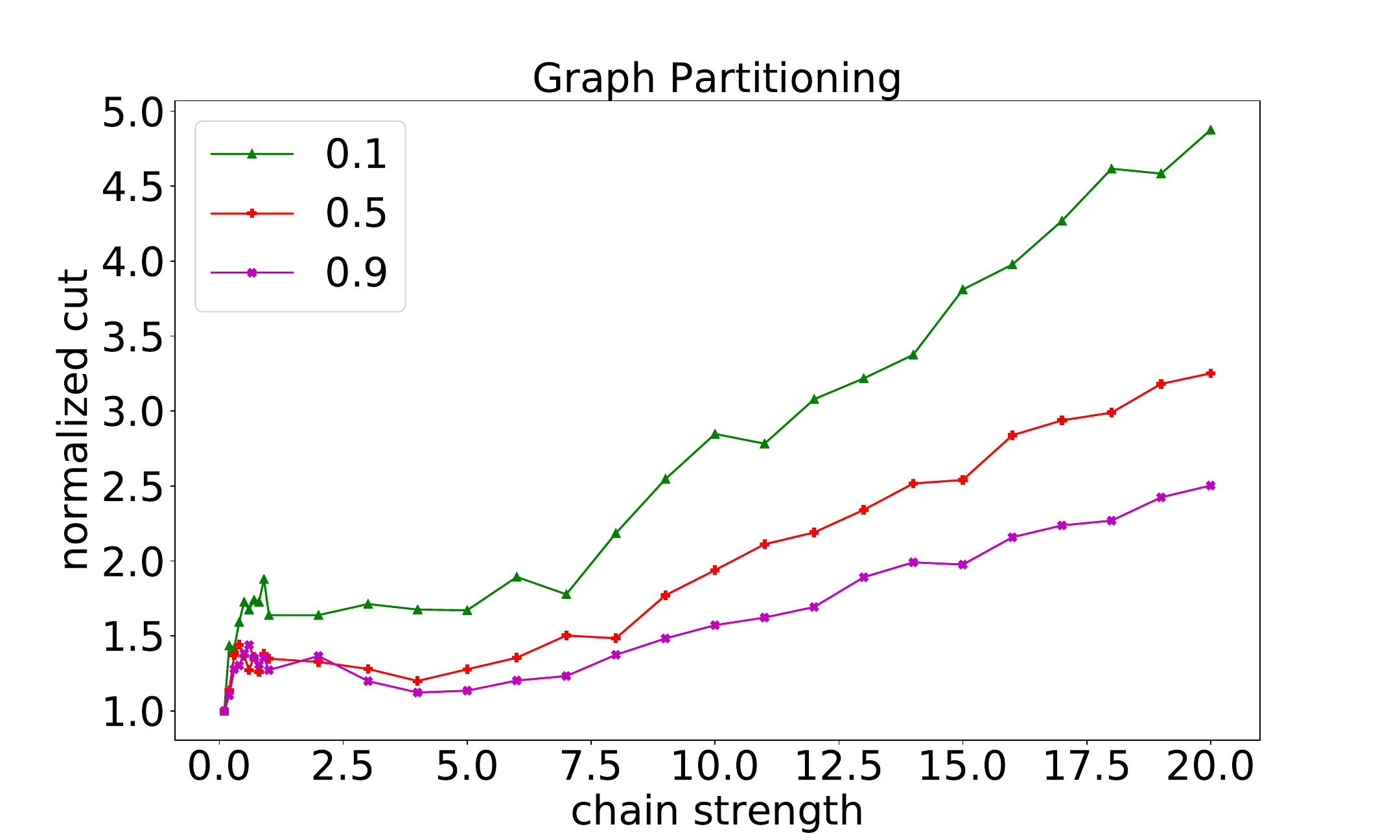}
    \caption{Normalized clique size for Maximum Clique (top left), cut size for Maximum Cut (top right), vertex cover size for Minimum Vertex Cover (bottom left), and cut size for Graph Partitioning (bottom right) as a function of the chain strength for three graph densities. The Graph Partitioning Ising  coefficients were scaled into range (-1, 1). \label{fig:chain_strength}}
\end{figure*}

Figure~\ref{fig:chain_strength} shows normalized results for the four problems as a function of the chain strength. The normalization has been done by dividing, for each graph problem and graph density, all of the objective function values by the absolute value of the minimum for that combination. This is done because the original objective function values have quite different ranges, thus making it hard to compare plots if the original values are used.

With the exception of the Maximum Clique problem, where the results are inconclusive, we observe that lower chain strengths tend to result in solutions of higher quality. Importantly, we see that non-standard chain strengths close to zero can lead to considerable performance improvements of our methods. This is due to the fact that in those cases, more chains are broken, thus giving our methods more room for optimization. Weighing the contributions from the D-Wave annealer and from a classical unembedding technique is a delicate balance, both for our methods and for D-Wave's unembedding techniques. For this reason we decided to employ the default \textit{uniform torque compensation} functionality provided by D-Wave to fix the chain strengths in our experiments of Section~\ref{sec:experiments}.